\documentclass[conference]{IEEEtran}
\IEEEoverridecommandlockouts
\usepackage{cite}
\usepackage{amsmath,amssymb,amsfonts}
\usepackage{algorithmic}
\usepackage{graphicx}
\usepackage{textcomp}
\usepackage{xcolor}
\usepackage[normalem]{ulem}
\def\BibTeX{{\rm B\kern-.05em{\sc i\kern-.025em b}\kern-.08em
    T\kern-.1667em\lower.7ex\hbox{E}\kern-.125emX}}

\usepackage{fancyhdr}
\begin{document}

\title{Distributed Attacks over Federated Reinforcement Learning-enabled Cell Sleep Control\\
{
}
}
\author{
    \IEEEauthorblockN{Han Zhang\IEEEauthorrefmark{1}, Hao Zhou\IEEEauthorrefmark{1}, Medhat Elsayed\IEEEauthorrefmark{2},Majid Bavand\IEEEauthorrefmark{2}, Raimundas Gaigalas\IEEEauthorrefmark{2},} \IEEEauthorblockN{Yigit
    Ozcan\IEEEauthorrefmark{2} and Melike Erol-Kantarci\IEEEauthorrefmark{1},\IEEEmembership{ Senior Member, IEEE}}
    
    \IEEEauthorblockA{\IEEEauthorrefmark{1} School of Electrical Engineering and Computer Science, University of Ottawa, Ottawa, Canada}
    
    \IEEEauthorblockA{\IEEEauthorrefmark{2} Ericsson Inc., Ottawa, Canada}
    
    \IEEEauthorblockA{\{hzhan363, hzhou098, melike.erolkantarci\}@uottawa.ca, }\IEEEauthorblockA{\{medhat.elsayed, majid.bavand, raimundas.gaigalas, yigit.ozcan\}@ericsson.com}
}

\maketitle

\thispagestyle{fancy}   
\fancyhead{}                
\lhead{Accepted by 2023 IEEE Globecom Workshops (GC Wkshps), \copyright2023 IEEE}
\cfoot{}
\renewcommand{\headrulewidth}{0pt}   
\begin{abstract}
Federated learning (FL) is particularly useful in wireless networks due to its distributed implementation and privacy-preserving features. However, as a distributed learning system, FL can be vulnerable to malicious attacks from both internal and external sources. Our work aims to investigate the attack models in a FL-enabled wireless networks. Specifically, we consider a cell sleep control scenario, and apply  federated reinforcement learning to improve energy-efficiency. We design three attacks, namely free rider attacks, Byzantine data poisoning attacks and backdoor attacks. The simulation results show that the designed attacks can degrade the network performance and lead to lower energy-efficiency.  Moreover, we also explore possible ways to mitigate the above attacks. We design a defense model called refined-Krum to defend against attacks by enabling a secure aggregation on the global server. The proposed refined-Krum scheme outperforms the existing Krum scheme and can effectively prevent wireless networks from malicious attacks, improving the system energy-efficiency performance.
\end{abstract}

\vspace{-5pt}
\section{Introduction} 
With the deployment of the 5G and beyond 5G (B5G) networks, the increasing traffic demand for cellular communications has reached an unprecedented level\cite{b5}. To meet diverse service requirements and facilitate intelligent wireless communications, various machine learning (ML) techniques have been used to solve problems in wireless networks\cite{b5-b}.

Reinforcement learning (RL) is a widely applied ML technique that provides automated solutions for high-complexity optimization problems\cite{b5-a}. 
Meanwhile, federated learning (FL) is another emerging ML technique that enables collaborative learning with local training in distributed systems, without sharing data. Federated reinforcement learning (FRL) is proposed as a combination of FL and RL and has proven effective in many wireless communication scenarios. For example, in \cite{b6}, FRL is used to allocate power resources and radio resources in network slicing. However, these achievements of using FRL are mainly accomplished in fully secure environments without considering malicious attacks.

Due to the inherently distributed implementation, FL is more vulnerable to malicious attacks than other centralized ML techniques. Distributed participants in FL are easier to be attacked and manipulated, and the parameter sharing and updating between local and global servers may expose the FL to potential risks \cite{b9}. 
As a result, it is crucial to investigate security issues in FL.

There are some existing studies about attacks and defenses for FL algorithms\cite{b10}. However, most research focuses on supervised learning and cannot apply to FRL models. 
In this work, we study the security problem in an FRL-enabled cell sleep control scenario. As the traffic load grows, 
improving network energy-efficiency and reducing energy costs become critical goals for wireless networks\cite{b4-a}. 
Performing sleep control to base stations (BS) to reduce energy consumption is a feasible way to improve energy-efficiency and make networks sustainable\cite{b4}. However, attacks on cell sleep control may cause different levels of system performance degradation. For example, it may waste system energy by making BSs never sleep or produce low throughput by keeping BSs in sleep mode.

In this paper, we first design an FRL-based cell sleep control scenario and BSs will cooperatively learn sleep control strategies through FL. Then we assume some BSs are malicious participants. Specifically, we propose three attack models, namely free rider, Byzantine data poisoning, and backdoor attacks specifically for the given cell sleep control scenario. To the best of our knowledge, this is the first work that applies the backdoor attacks to a wireless network control application. The simulation results show that the designed attacks will lower system energy-efficiency. Meanwhile, we also propose a defense scheme called refined-Krum to defend against these attacks. Compared with the existing Krum defense scheme, it can achieve a better defense effect without knowing the number of attackers.  

The rest of the paper is organized as follows. Section \ref{s2} introduces related works, and Section \ref{s3} shows our system model. 
Section \ref{s4} introduces FRL-based sleep control scenario, 
and Section \ref{s5} presents the designed attacks and the proposed defense model. Finally, Section \ref{s6} shows simulation results, and Section \ref{s7} concludes this work.
\vspace{-5pt}
\section{Related Works}
\label{s2}
There have been many studies that design attacks and defenses towards breaches in FL algorithms. In \cite{b11}, data poisoning attacks are performed on FL-based image classification problems. \cite{b12} performs backdoor attacks on the FL system with single or multiple malicious participants. \cite{b13} proposes secure aggregation methods to defend Byzantine data poisoning attacks in the FL system. These works are only designed for supervised learning and do not apply to RL models. \cite{b14} and \cite{b15} proposes data poisoning attacks and defenses for FRL. However, these works are only tested with ready-to-use data sets and have some limitations if applied to complicated wireless network scenarios. 

Meanwhile, other works study attacks on FL in wireless networks. \cite{b15-1} designs attacks specific to the wireless traffic prediction models in centralized and distributed scenarios. However, this work also uses a supervised learning model, and its attack method cannot be directly applied to other wireless network control applications that typically use RL. In \cite{b15-2}, over-the-air jamming attacks on the uplink and downlink of FL in wireless networks are studied. But it only focuses on external attacks and fails to study the internal attacks in FL.

There are also some studies on cell sleep control for energy saving. \cite{b4} improves energy-efficiency of small cell networks by switching BSs to different modes. In \cite{b16}, an RL-based traffic adaptive sleep mode control algorithm for BSs is proposed. However, these works are accomplished in fully secure environments and fail to consider attacks and defense. Different from existing studies, our work designs attacks to the specific FRL-based cell sleep control scenario. We focus more on vulnerabilities of FRL models related to realistic wireless environments and evaluate the effectiveness of attacks based on wireless network performance metrics.

\section{System Model}
\label{s3}
\begin{figure}[!t]
\centerline{\includegraphics[width=2.9in]{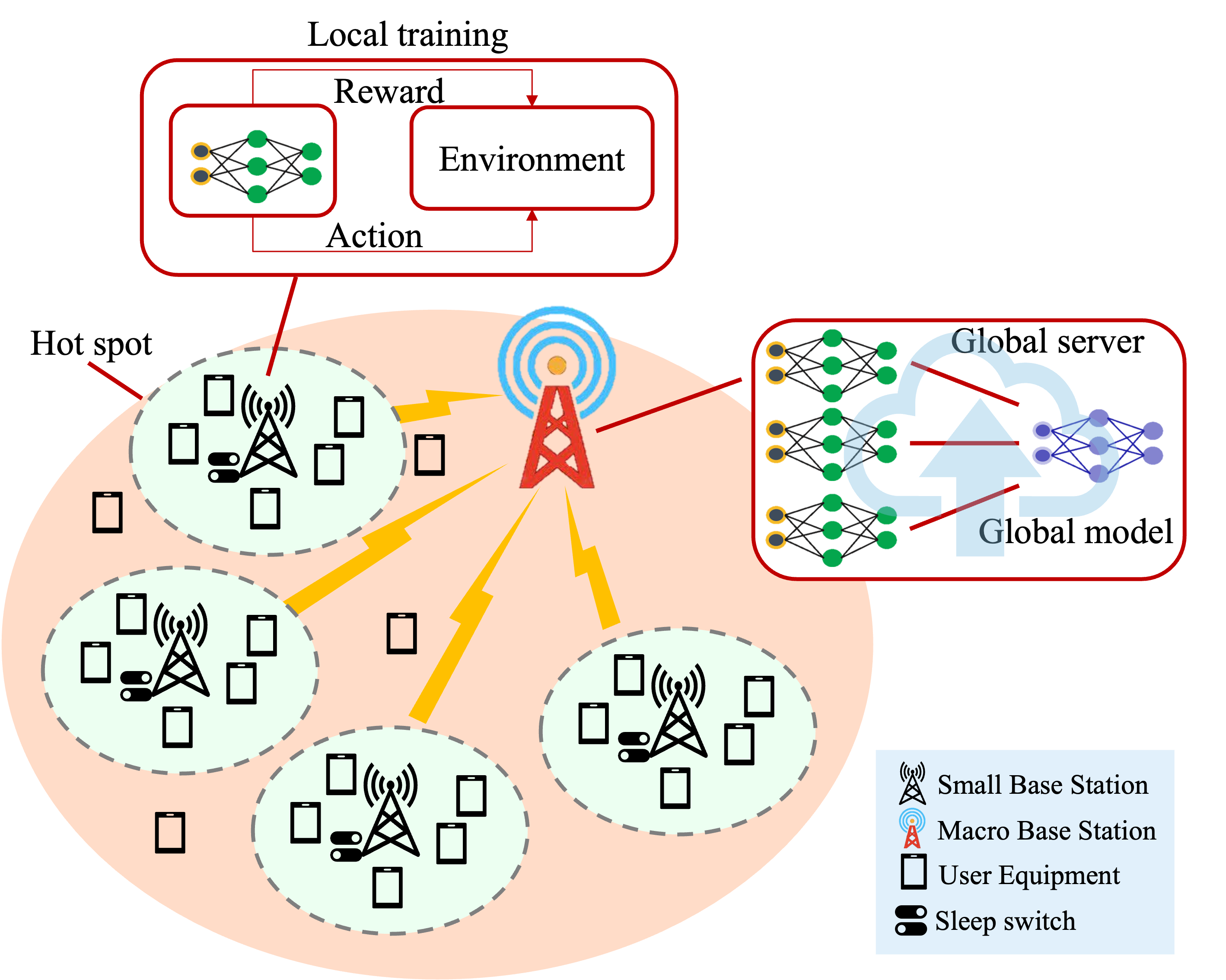}}
\caption{System Model.}
\label{fig1}
\vspace{-15pt}
\end{figure}

The system model is shown in Fig. \ref{fig1}. We  consider a heterogeneous cellular network consisting of multiple BSs. There is one macro BS (MBS) cell and $N$ small BS (SBS) cells cooperatively serving $M$ distributed user equipment (UE) and handling traffic loads. 
The MBS is always active to ensure coverage and is responsible for controlling data services. 

To effectively save energy costs of the system, we adopt three different sleeping modes for SBS cells, which are active, sleep, and deep sleep\cite{b4}. Active means SBSs are in full operation and consume the most energy. Sleep means SBSs temporarily stop transmitting data for the UEs but can be easily waken up and decide whether to continue sleeping in the next iteration. Deep sleep means more components are deactivated to save more energy, and SBSs take longer time to wake up. If a SBS turns to sleep, the arriving traffic will be offloaded to the MBS. SBSs sleeping at inconvenient intervals can cause low energy efficiency or data congestion in MBS traffic, thus degrading system performance.

This scenario considers a downlink orthogonal frequency-division multiplexing cellular system. The link capacity between the $m^{th}$ UE and the $n^{th}$ SBS can be given as follows:
\begin{equation}
C_{n,m} = \delta_{n}\sum_{r\in R_{n}} B_{r} log_{2}(1+SINR_{n,m,r}),\label{eq1}
\end{equation} 
where $\delta_{n}$ is a binary indicator to denote whether the $n^{th}$ SBS is active or sleeping. $R_{n}$ denotes the set of available resource blocks of the $n^{th}$ SBS and $B_{r}$ denotes the bandwidth of the $r^{th}$ resource block. $SINR_{n,m,r}$ denotes the signal to interference noise ratio (SINR) between the $m^{th}$ UE and the $n^{th}$ SBS on the $r^{th}$ resource block, which can be given as:
\begin{equation}
SINR_{n,m,r} =\frac{\beta_{n,m,r} g_{n,m} P^t_{n}}{\underset{n'\in N, n'\neq n}{\sum}\ \underset{m'\in M_{n'}}{\sum} \beta_{n',m',r} g_{n',m} P_{n'}+B_{r}N_{0}},\label{eq2}
\end{equation} 
where $\beta_{n,m,r}$ is a binary indicator to denote whether the $r^{th}$ resource block of the $n^{th}$ SBS is allocated to the $m^{th}$ UE. $g_{n,m}$ is the channel gain of the transmission link, which is decided by a free space propagation model. $P^t_{n}$ denotes the transmission power of the $n^{th}$ SBS and $N_{0}$ denotes the noise power density. 

We assume that UEs can support dual connectivity and can simultaneously connect to the MBS and SBS\cite{b17-1}. If the SBS is active, the UE will be served by the SBS. Otherwise, it will be served by the MBS. The energy-efficiency of the system can be defined as:
\begin{equation}
EE = \frac{\sum_{m\in M} b_m}{\sum_{n\in N}P_n + P_0},\label{eq3}
\end{equation}
where $b_m$ denotes the throughput of the $m^{th}$ UE, which is decided by both link capacity and arriving traffic. $P_0$ and $P_n$denotes the power consumption of the MBS and the $n^{th}$ SBS. 

The optimization objective of the sleep control application is to achieve high energy-efficiency. Here we formulate the problem as: 
\begin{align}
 \underset{a_n}{max}\ &EE-\sum_{m \in M}\epsilon_m, \label{eq4}\\
s.t.\ &(\ref{eq1})-(\ref{eq3})\nonumber
\\& a_n \in \{0,1,2\},\ \forall n \in N \tag{4a}\label{eq4a}
\\& \delta_n = \begin{cases}
= 1, &if\ a_n = 0,\\
= 0, &else
\end{cases}\tag{4b}\label{eq4b}
\\& P_n = \begin{cases}
P_w, &if\ a_n = 0,\\
0.5P_w, &if\ a_n = 1,\\
0.35P_w,&else\\
\end{cases}\tag{4c}\label{eq4c}
\end{align}
where $\epsilon_m$ denotes the packet drop rate of the $m^{th}$ UE. A packet will be dropped if it exceeds the transmission delay constraint\cite{b17-a}. $a_n$ denotes the sleeping modes of the $n^{th}$ SBS. $a_n=0$ indicates the SBS is in the active mode, $a_n=1$ indicates the SBS is in the sleep mode, and $a_n=2$ indicates the SBS is in deep sleep mode. $P_w$ denotes the energy consumption of the SBS in active mode. The sleep mode can reduce energy consumption by 50\%, and the deep sleep mode can reduce it by 65\%\cite{b4}. 

To solve this problem, we use federated reinforcement learning (FRL) to promote privacy-preserving collaborative training. Each SBS holds a local deep reinforcement learning (DRL) model, which observes states and rewards from the environment and selects actions by choosing an adequate sleeping mode. The MBS serves as a global server in FRL, collecting local models from SBSs for model aggregation and distributing the global model as feedback. To attack the system, we suppose $N^{mali}$ out of the $N$ SBSs are malicious and can cause system performance degradation by updating malicious local models to the global server.

\section{Federated reinforcement learning-based cell sleep control}
\label{s4}
This section introduces the FRL-based cell sleep control application. Here DRL is applied in each SBS as a local model, and the optimal actions are selected by maximizing the long-term expected rewards.

The Markov decision process (MDP) of each local DRL is defined as follows:
\begin{itemize}
    \item State: The state includes the sleeping mode of the SBS and the traffic load of the SBS and the MBS in the past 5 transmission time intervals
    which can be used to estimate the upcoming traffic load. It also includes the current delay and throughput of the SBS, which can be given as:
    \begin{equation}
     s_n = \{\delta_n, L_n, L_0, d_n, b_n\},\forall n \in N,\label{eq6}
     \end{equation}
     where $L_n$ denotes the traffic load of the $n_{th}$ SBS. $L_0$ denotes the traffic load of the MBS. $d_n$ and $b_n$ denote the delay and the throughput.
     \item Action: The action of sleep control is to choose an adequate sleeping mode for the SBS, which can be given as:
     \begin{equation}
     a_n = \{0,1,2\},\forall n \in N,\label{eq7}
     \end{equation}
     \item Reward: The reward function is defined as a combination of both quality of service (QoS) related indicators and the power consumption related cost, which can be given as:
     \begin{equation}
     R_n = \eta_1 b_n - \eta_2\epsilon_n  - \eta_3 P_n,\forall n \in N,\label{eq8}
     \end{equation}
     where $\epsilon_n$ denotes the packet drop rate and $b_n$ denotes the throughput. $\eta_1$, $\eta_2$ and $\eta_3$ are the coefficients used to balance different rewards. When obtaining a high reward value, we expect the system to consume as little energy as possible while ensuring a high throughput. Therefore, maximizing the given reward value is equivalent to maximizing the energy-efficiency and minimizing the packet drop rate.
\end{itemize}

On top of local models, we apply FRL to enable collaborative training and accelerate learning while keeping data locally and preserving privacy. In each FRL cycle, the local models will first perform local training according to local experience, which can be given as:
\begin{equation}
\begin{split}
\theta_n^{t+1} &= \theta_n^t + \alpha[r_n^t + \gamma \mathop{max}\limits_{a}Q(s_n^{t+1},a;\theta_n^t) \\ & -Q(s_n^t,a_n^t;\theta_n^t)]\nabla Q(s_n^t,a_n^t;\theta_n^t),\label{eq9}
\end{split}
\end{equation}
where $\theta_n$ denotes the local model parameters of the $n_{th}$ SBS, $\alpha$ denotes the learning rate and $\gamma$ denotes the discount factor. $Q(s_n^t,a_n^t;\theta_n^t)$ denotes the long-term expected reward of the $n_{th}$ SBS choosing the action $a_n^t$ under the state $s_n^t$.
After local training, the local models are uploaded to the global server for model aggregation, which can be formulated as:
\begin{equation}
\theta_G^{t+1} = \sum^{n=1}_{N}w_n\theta_n^{t+1}
\end{equation}
where $\theta_G$ is the parameters of the global model. $w_n$ is the weight of the $n_{th}$ local model and it is decided by the number of training samples. In the scenario of FRL-based cell sleep control, we assume all the local models are equally weighted.

After the global model aggregation, the global model parameters are sent back to the SBSs and the local models are updated by replacing the local parameters with global parameters.

\vspace{-10pt}
\section{Attacks and Defense}
\label{s5}
\begin{figure}[!t]
\centerline{\includegraphics[width=3.0in]{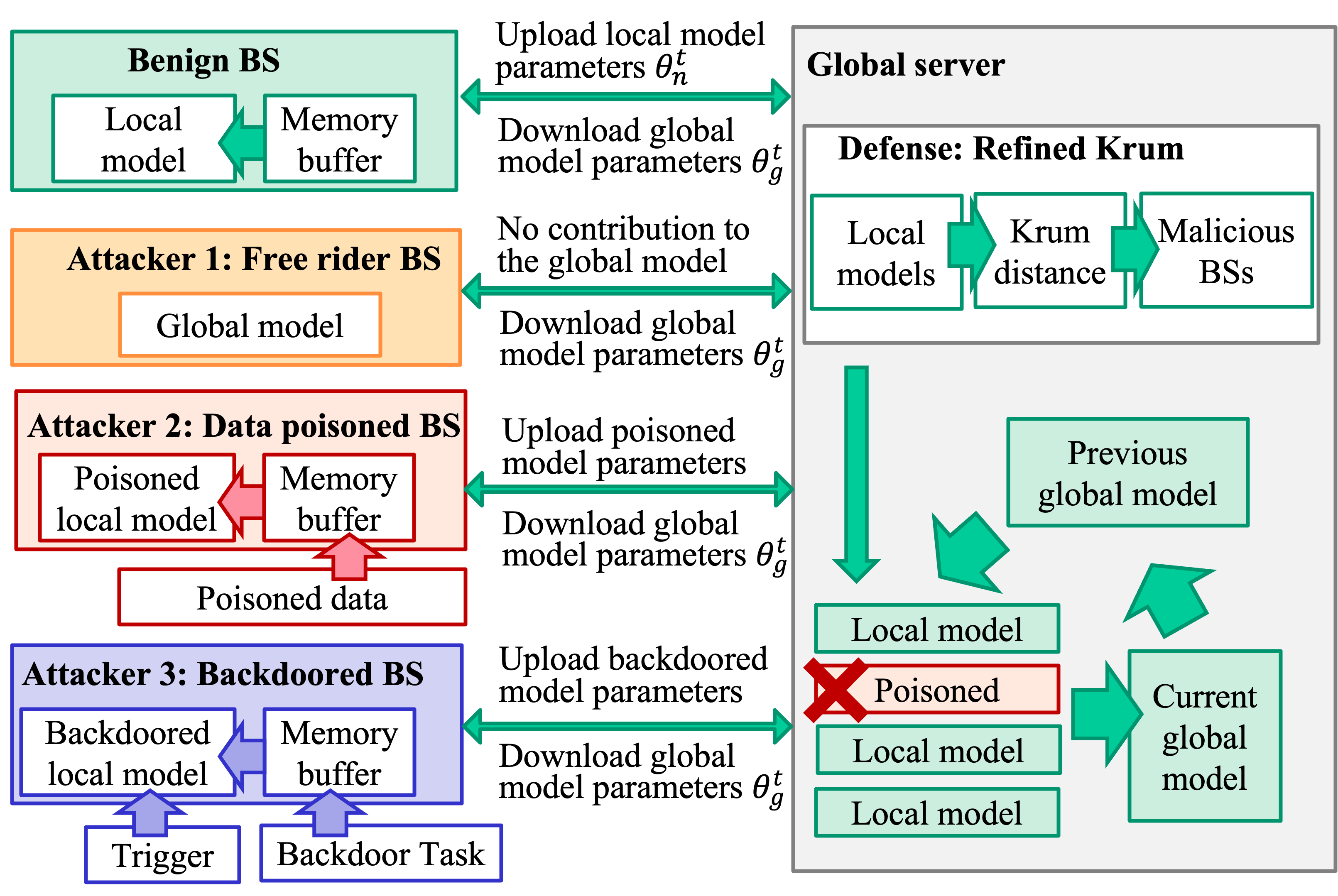}}
\caption{Attack and defense models.}
\label{fig2}
\vspace{-15pt}
\end{figure}

This section presents the designed attack and defense models in the FRL-based cell sleep control scenario.
Fig. \ref{fig2} shows the structure of the investigated attacks and the proposed defense model. We proposed three attack models: free rider attacks, Byzantine data poisoning attacks, and backdoor attacks. We also proposed one defense scheme called refined-Krum.

\subsection{Attack models.}
\subsubsection{Free rider attacks.}
Free riders refer to the FL participants who do not train their local models during the local training step \cite{b19}. As shown in Fig. \ref{fig2}, a benign BS will keep a memory buffer to store local experience and use it to train a benign local model. In contrast, a free rider does not 
train its local model and will submit the previously received global model as its own local model.
The free rider is a passive attack method which means it will not actively destroy the global model or other local models. However, free riders are still malicious because they enjoy the resources and efforts of collaborators without contributing their own experience and training results during the collaboration. In this way, they can break the fairness of the FL system, and when the proportion of free riders increases, they will slow down the FL training.

\subsubsection{Byzantine data poisoning attacks.}
Byzantine data poisoning attacks are implemented by injecting poisoned data into the local memory buffer of the malicious participants and making participants train their local model in the wrong direction \cite{b11}. As shown in Fig. \ref{fig2}, the poisoned local model is trained with a mixture of local experience and poisoned data. 

For our FRL model, 
we create poisoning data by giving the model a larger reward if the SBS chooses to be active and a smaller reward if the SBS chooses to sleep. The local model training in Eq. (\ref{eq9}) on the poisoned data can be reformulated as:
\begin{equation}
\begin{split}
\theta_n^{t+1} &= \theta_n^t + \alpha[r_n^t +\sigma R^{ex}+ \gamma \mathop{max}\limits_{a}Q(s_n^{t+1},a;\theta_n^t) \\ & -Q(s_n^t,a_n^t;\theta_n^t)]\nabla Q(s_n^t,a_n^t;\theta_n^t),\label{eq10}
\end{split}
\end{equation}
where $R_{ex}$ denotes the extra reward and $\sigma$ is an indicator to denote whether to add an extra reward or to reduce the reward. If $a_n^t = 0$, $\sigma = -1$. Otherwise, 
$\sigma = 1$. 

After local training, the poisoned local model will be uploaded to the global server and the effect of poisoned data will then be propagated to other local models during the global model aggregation. 

\subsubsection{Backdoor attacks.}

Different from regular data poisoning attacks, backdoor attacks define backdoor tasks to make a separation from the main tasks. Main tasks refer to the task defined by the original training data.
Backdoor tasks are defined by the attacker, and they represent a special input data pattern that rarely appears in the original training data. In FRL, the backdoor tasks can be defined by specific states. Every time the backdoor attackers want to attack the system model, they can trigger the backdoor pattern and make the model select abnormal actions.

In our sleep control scenario, we define the backdoor task as an SBS with a very high traffic load. It can be easily triggered by putting a UE with an extremely heavy traffic load into the system, and the trigger result is defined as the SBS will turn into deep sleep mode. As shown in \ref{fig2}, the backdoor-ed local model is trained with the mixture of local experience data and synthetic backdoor task training data. 

Benefiting from separating the main and backdoor tasks, the backdoor attacks are stealthy and difficult to detect. 
Since the backdoor pattern is defined by the attacker and unknown to others, the model owner will only test it with main tasks while doing a security check. In this way, they could not find the model had already been backdoor-ed. 

\subsection{Defense model.}
We propose a refined-Krum defense model based on the existing secure FL aggregation method Krum\cite{b13}. As shown in Fig. \ref{fig2}, the refined-Krum is deployed at the global server and will be performed before global aggregation during each FL iteration. In this subsection, we first introduce the Krum defense scheme and then illustrate how the refined-Krum model is defined. 
\subsubsection{Krum defense}
Krum is proposed in \cite{b13} and its core idea is to assume that all benign local models are similar. Therefore, the malicious models can be found by measuring the similarity of all the local models by the Krum distance. 

In the Krum defense, the Krum distance for each local model is first calculated. In the first step, the Euclidean distance between parameters of the $n^{th}$ local model and the global model in the last FL iteration is calculated as:

\begin{equation}
G^{t+1}_n = \left\|\theta^{t+1}_n-\theta^{t}_G\right\|_2\label{eq11}
\end{equation}

Then, the distance between the $n^{th}$ local model and $k^{th}$ local model can be given as:

\begin{equation}
D^{t+1}_{nk} = \left\|G^{t+1}_n-G^{t+1}_k\right\|_2\label{eq12}
\end{equation}

The distance between each local model and all other local models is then added. In this way, the Krum distance for each local model can be obtained, which can be given as:

\begin{equation}
D^{t+1}_{n} = \sum_{k \in N}D^{t+1}_{nk}\label{eq13}
\end{equation}

Finally, the Krum defense will select the local model with the smallest Krum distance and replace the global model with the selected local model.

\subsubsection{Refined-Krum}
Although the Krum defense scheme is proven to be effective in some cases, choosing only a local model for global aggregation is quite unstable and it may not get the full benefit of FL. Therefore, we designed a new defense algorithm called refined-Krum. It can be concluded into four steps, which are calculating the similarity gaps, estimating the number of malicious participants, identifying malicious participants and secure aggregation. 
\begin{itemize}
    \item Calculating the similarity gaps. In the first step of refined-Krum, we calculate the Krum distance of each local model to evaluate the similarities between models. But instead of only selecting the local model with the smallest Krum distance, we sort all the models by their Krum distances from the smallest to the largest and calculate the gap between two adjacent Krum distances.
    \item Estimating the number of malicious participants. With the gap values calculated in the first step, we can then estimate the number of malicious participants by finding the maximum gap between the given distance list. This is based on the assumption that most models are benign and the malicious models are quite different from the benign ones. So there will be a large gap between the similarities. 
    \item Identifying malicious participants. If the maximum gap is much larger than the average, we suppose it could precisely separate malicious models from benign ones and decide the threshold for the Krum distance. If the Krum distance of the $n^{th}$ local model is larger than the threshold value, the $n^{th}$ SBS is treated as a malicious participant. Other SBSs are treated as benign participants. On the other hand, if the maximum gap is close to the average gap, we assume that the threshold cannot be accurately determined and to mitigate the risk of being attacked, only one benign model will be selected.
    \item Secure aggregation. After identifying the malicious participants, we can perform secure aggregation by admitting only benign models into the global model aggregation. This prevents the malicious data from influencing the global model and the benign SBSs. At the same time, we will make MBS take over the sleep control for the malicious SBSs and prevent attackers from controlling these SBSs.
    
    
\end{itemize}
\vspace{-5pt}
\section{Numeric results}
\label{s6}
\subsection{Simulation settings.}
In the simulation, we consider 8 SBSs, and each SBS has 10 UEs. The fixed power consumption of MBS and SBSs are 40W and 20W, respectively\cite{b17-1}. The radius of MBS and SBSs are 400m and 100m, respectively. The available bandwidth for each SBS is 20 MHz, and for MBS is 10 MHz. $\eta_1$, $\eta_1$ and $\eta_1$ are respectively 0.1, 1 and 0.01. During the simulation, we change the average traffic load of each SBS from 30 Mbps to 70 Mbps and compare the system energy-efficiency under different attack and defense models. We simulate a 24-hour typical residential area traffic pattern in each TTI, which is given from \cite{b18}.


For the free rider attacks, we include two free riders in the network. For Byzantine data poisoning attacks and backdoor attacks, we have one malicious SBS in the networks. We assume the proportion of poisoned data or backdoor task training data of malicious SBSs is 5\%. Therefore, the proportion of poisoned data in the total data of all the SBSs is 0.625\%.

\subsection{Simulation results}
\begin{figure}[t]
\centerline{\includegraphics[width=2.9in]{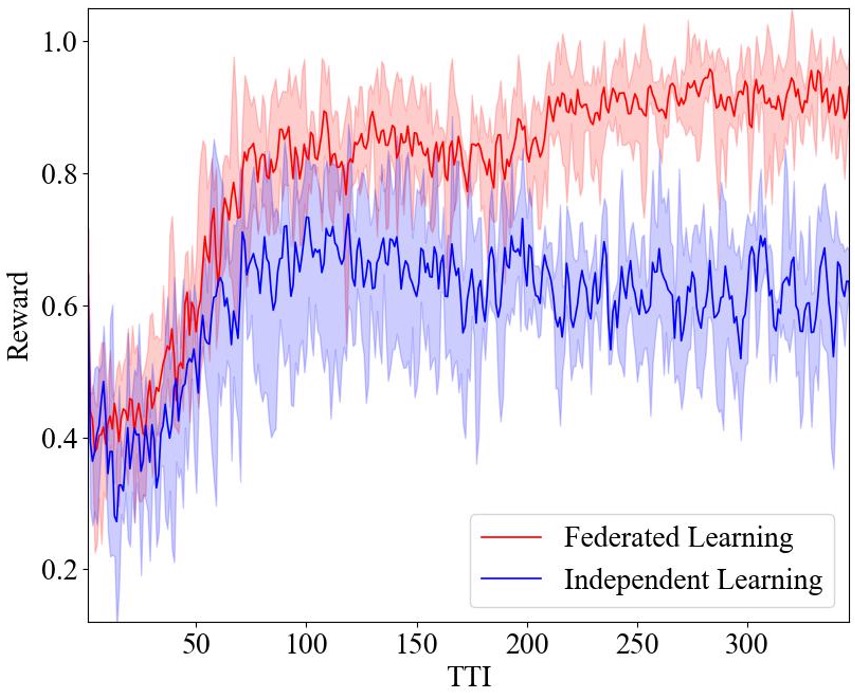}}
\vspace{-5pt}
\caption{The convergence curves of FL and independent learning.}
\label{fig4}
\vspace{-10pt}
\end{figure}

Firstly, we compare the simulation results of FL and independent learning-based cell sleep control in a fully secure environment. Independent learning (IL) means there are no collaborations between SBSs, and each SBS will train a DRL model according to their local buffer data. Fig. \ref{fig4} shows the convergence curves of FL and independent learning when the average traffic load of each SBS is 40 Mbps and in each TTI we run 24-hour traffic. The FL algorithm we use during the simulations is FedAvg. The FL performs much better than IL and has higher rewards, which demonstrates the effectiveness of the FL algorithm.

\begin{figure}[t]
\centerline{\includegraphics[width=2.9in]{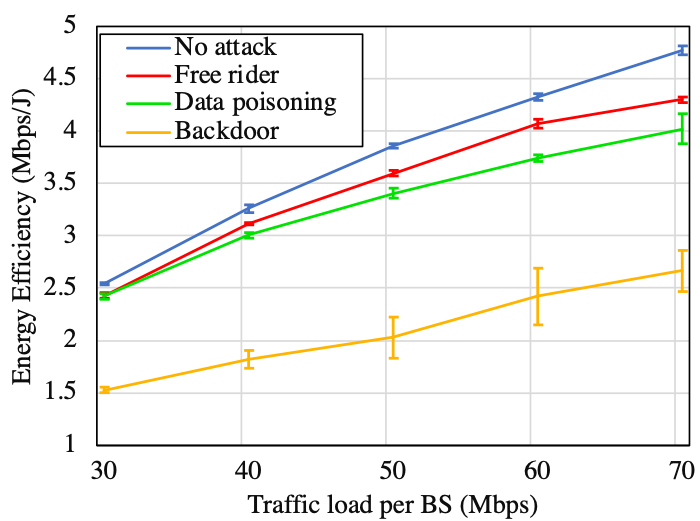}}
\vspace{-5pt}
\caption{The system energy-efficiency under different attacks.}
\label{fig6}
\vspace{-15pt}
\end{figure}

Then we add three different attacks to the FRL-based sleep control scenario. The system energy-efficiency under free rider attacks, data poisoning attacks, and backdoor attacks are shown in Fig. \ref{fig6}. The system performance in a secure environment with no attacks is also compared. It can be observed that three kinds of attacks can degrade the system performance to different levels. Backdoor attacks can be seen as the most effective attacker. When the average traffic load is 70 Mbps, the backdoor attacks can reduce the system energy-efficiency by 52\%. Among the remaining two attacks, the data poisoning attacks are more effective than the free rider attacks, even with fewer attackers. From this observation, we can also conclude that while designing defense mechanisms for FRL, it is more important to prevent malicious participants from being involved in the aggregation than to ensure that the benign participants are involved in the aggregation. When the average traffic load is 70 Mbps, a malicious SBS with poisoning attacks can reduce the system energy-efficiency by 18\%.

\begin{figure}[t]
\centerline{\includegraphics[width=2.8in]{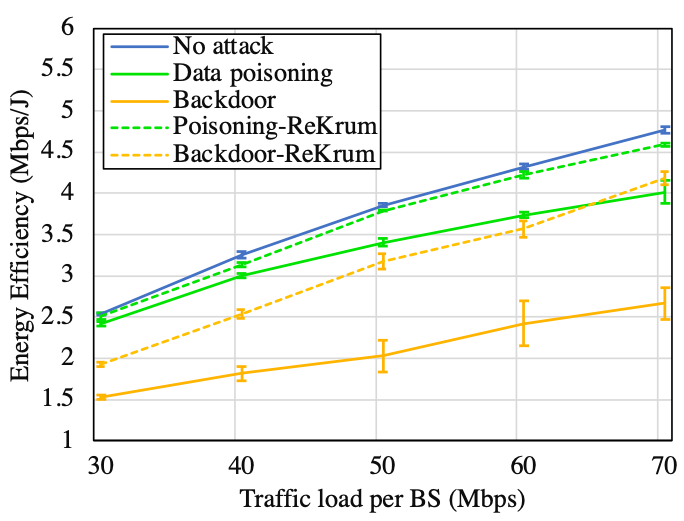}}
\vspace{-5pt}
\caption{The system energy-efficiency under the proposed defense scheme.}
\label{fig8}
\vspace{-10pt}
\end{figure}

When it comes to defense, the system energy-efficiency under our proposed refined-Krum defense scheme is shown in Fig. \ref{fig8}. We only defend against poisoning and backdoor attacks because even if we can detect free riders, we cannot force them to contribute to the model. As it can be observed, for data poisoning attacks, the system energy-efficiency after the defense is very close to the situations with no attack. We can conclude that the defense can almost fully recover the system from data poisoning attacks with a limited number of attackers. The defense scheme can also significantly improve system performance and increase energy-efficiency for backdoor attacks. However, the energy-efficiency defense is still lower than the performance in a secure environment. This indicates that our proposed defense scheme is quite effective for some attacks but less effective for others.

\begin{figure}[t]
\centerline{\includegraphics[width=2.8in]{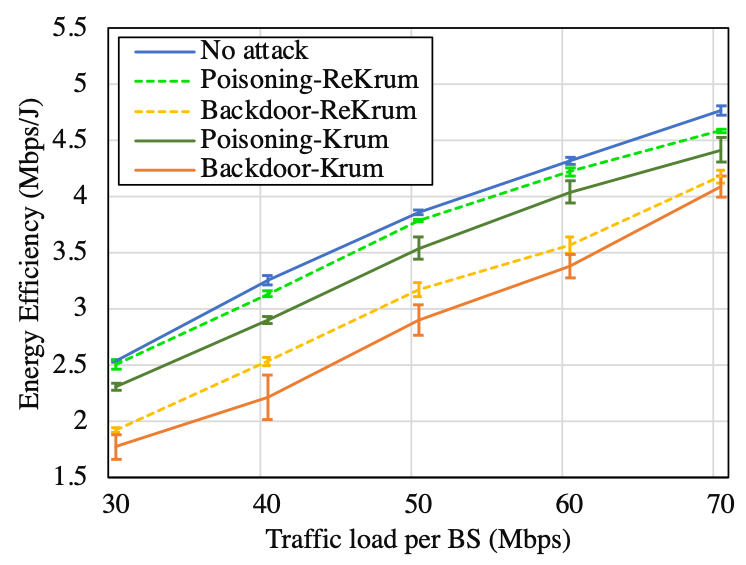}}
\vspace{-5pt}
\caption{The system energy-efficiency under the Krum defense scheme and the proposed defense scheme.}
\label{fig9}
\vspace{-20pt}
\end{figure}

In Fig. \ref{fig9}, we further compare our proposed refined-Krum defense scheme with the existing Krum defense scheme. For both kinds of attacks, our proposed refined-Krum defense scheme can get a higher energy-efficiency compared with the Krum defense scheme. Also, refined-Krum is more stable with a smaller confidence interval.

\vspace{-5pt}
\section{Conclusion}
\label{s7}
In this work, we studied how to attack a FRL-based cell sleep control scenario in a wireless network. We considered three types of attacks that could perform on wireless networks, which are free riders, Byzantine data poisoning attacks and backdoor attacks. According to the simulation results, these attacks can degrade system performance with lower energy-efficiency. We also proposed a defense scheme called refined-Krum to defend against these attacks. The simulation results show that our proposed defense scheme can effectively increase the system energy-efficiency and prevent the system from attacks. 
In our future research, we plan to investigate more advanced attacks and improved defense schemes. 
\vspace{-5pt}
\section*{Acknowledgement}
 This work has been supported by MITACS and Ericsson Canada, and NSERC Collaborative Research and Training Experience Program (CREATE) under Grant 497981.

\end{document}